\documentclass[twocolumn,showpacs,preprintnumbers,amsmath,amssymb]{revtex4}
\usepackage{graphicx}
\usepackage{dcolumn}
\usepackage{bm}

\begin{document}


\title{Interface effects in ferroelectric PbTiO$_3$ ultrathin films on a
paraelectric substrate}

\author{M. Sepliarsky}
\affiliation {Instituto de F\'{\i}sica Rosario, 
       Universidad Nacional de Rosario. \\
       27 de Febrero 210 Bis,  2000 Rosario, Argentina}


\author{M.G. Stachiotti}
\affiliation {Instituto de F\'{\i}sica Rosario, 
       Universidad Nacional de Rosario. \\
       27 de Febrero 210 Bis,  2000 Rosario, Argentina}
	       
\author{R.L. Migoni}
\affiliation {Instituto de F\'{\i}sica Rosario, 
       Universidad Nacional de Rosario. \\
       27 de Febrero 210 Bis,  2000 Rosario, Argentina}
		      
\date{\today}

\begin{abstract}
Interface effects on the ferroelectric behavior of PbTiO$_3$ ultrathin
films deposited on SrTiO$_3$ substrate are investigated using an interatomic
potential approach with parameters fitted to first-principles
calculations. We find that the correlation of
atomic displacements across the film/substrate interface is crucial for the stabilization of the ferroelectric state in films a few unit-cells thick. 
We show that the minimum film thickness for the appearance of a spontaneous polarized domain state is not an intrinsic property of the ferroelectric film but depends on the polarizability of the paraelectric substrate.
We also observe that the substrate displays an induced polarization 
with an unusual oscillatory behavior.

\end{abstract}

\maketitle

The current high interest in ferroelectric thin films and nanostructures
is due to the wide range of their potential applications in
microelectronics. 
It is well known that the reduction of the
characteristic size of ferroelectric structures in the nanoscopic regime
may result in finite size effects. For instance, whether or not there is
a critical thickness below which no ferroelectricity can occur is still
under debate. Ferroelectricity is a cooperative phenomenon and, as such, it is
expected to be strongly influenced by surfaces. 
Most active device applications eventually
involve thin films forms and the understanding of the near-surface
region becomes important. 
A ferroelectric thin film cannot be considered in 
isolation, but rather the measured properties reflect the entire system 
of films, interfaces, electrodes, and substrates. 
Although the effects of epitaxial strain and metal electrodes on 
film properties have been extensively investigated ~\cite{dawb05}, 
almost nothing is known on how a paraelectric substrate modifies 
the ferroelectric properties of thin films.

Recently, Fong et. al. showed that ferroelectric phases can be
stable down to $\approx 12 \AA$ (three unit cells) in PbTiO$_3$ films
deposited on SrTiO$_3$  substrates  by forming 180$^o$ stripe domains,
suggesting that no fundamental thickness limit is imposed by intrinsic
size effects in thin films~\cite{fong04}.  This was directly confirmed by
x-ray photoelectron diffraction studies~\cite{desp05}. 
Moreover, high-resolution electron density maps obtained by
synchrotron x-ray scattering showed that thin ferroelectric films grown
epitaxially on paraelectric substrates display a rich variety of
structures and properties. It was shown that monodomain
ferroelectricity can be stable when the samples are cooled sufficiently
slowly, and details of the PbTiO$_3$/SrTiO$_3$ interface were 
obtained~\cite{fong05}. 
As for the origin of ferroelectricity in such ultra-thin films,
a mixed Pb$_x$Sr$_{1-x}$TiO$_3$ interface has been suggested~\cite{glinc05}. 
Despite these efforts, it is not clear why a polar state
remains stable at such low thicknesses.    

In this work we investigate the effects of a SrTiO$_3$ substrate on
the ferroelectric behavior of PbTiO$_3$ thin films using an atomic level
description. We find that the minimum film thickness for
ferroelectricity 
is not an intrinsic property of the ferroelectric film
but depends on the polarizability of the paraelectric substrate. 
We show that the correlation of atomic displacements between 
film and substrate ions across the interface 
make possible 
the stabilization of the
ferroelectric state in films up to 2 unit cells thick.
We also find that a region of the SrTiO$_3$ substrate near the film
polarizes in the opposite direction than the film. 

The simulations are carried out using shell models with parameters
fitted to first-principles calculations. That is, no explicit
experimental data are used as input. The shell model offers a
reliable atomic level description of ferroelectric materials~\cite{sepli05},
and is a practical tool to investigate properties of systems where a
large number of atoms are involved. The particular model for PbTiO$_3$ 
(PT) used here reproduces correctly the cubic-tetragonal phase transition
of the bulk~\cite{sepli04}, and it is also able to
describe properly surface properties, such as atomic relaxation
patterns, change in the interlayer distances, layer rumpling, and the
antiferrodistortive surface reconstruction observed 
experimentally~\cite{sepli051}. 
The model for the SrTiO$_3$ (ST) substrate is developed following
a similar procedure; that is the parameters are fitted to
first-principles calculations. Moreover, to make the ST model compatible
with the PT model, the only difference between both lies in the
different A-Ti and A-O interactions and the different polarizability
parameters for Sr and Pb. The input data correspond to LDA
calculations of the energy as function of volume, lattice dynamics and
underlying potential energy surfaces as function of relevant
distortions. The resulting model displays a cubic lattice parameter
a=3.860$\AA$ that agrees with the LDA value of 3.861 $\AA$
(we note that LDA underestimates the lattice constant of perovskites
by $\approx$ 1$\%$, a better agreement with experiments is obtained 
with other functionals~\cite{heife01}).  
The ferroelectric soft-mode mode is stable (77 cm$^{-1}$) at the
theoretical equilibrium volume and unstable (i113 cm$^{-1}$) at the
experimental one, in agreement with the LDA behavior (68 cm$^{-1}$ and i94
cm$^{-1}$ at the respective volumes). 

We simulate coherently strained PT films on (001) ST substrates using a
semi-infinite crystal geometry. The z-axis lies along the growth
direction of the film, and the x and y axes are chosen to be along the
pseudocubic $[$100$]$ and $[$010$]$ directions.  The simulation cell
contains $10 \times 10$ unit cells with periodic boundary conditions in the x-y
plane. The free surface is at the top and corresponds to a
PbO-terminated surface, which is the equilibrium termination 
for PT~\cite{meyer99}. 
The interface between the film and the substrate is chosen to
be coherent, as it is observed in experiments~\cite{fong05}, 
and consist of a common
TiO$_2$ layer. We consider that a film with thickness of $n$ unit cells 
corresponds to $n$ PbO layers.
The film-substrate system is described with a
region I - region II strategy. Region I corresponds to the film and the
nearby part of the substrate in interaction with the film. The second
region corresponds to the part of the substrate which is away from the
interface, and it simulates the interior of the non disturbed crystal.
Whereas atoms in region I are allowed to relax, atoms in region II are
held fixed at the ideal cubic positions. For the substrate, a thickness
of 10 unit cells is taken for region I, which is thick enough to avoid
significant changes in the final results. The long range
electrostatic energy and forces are calculated by a direct sum method
~\cite{wolf99}. The equilibrated zero-temperature structures are determined by
standard atomic relaxation method until the force on each individual ion
is less than 0.001 eV/$\AA$.  

\begin{figure} [tbp] 
\includegraphics[height=3.5in,width=3.35in]{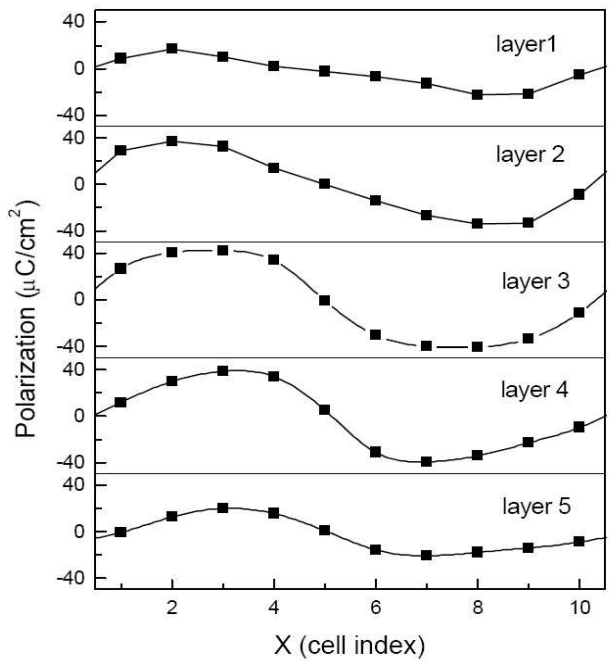}
\caption{\label{fig1}Cell-by-cell out-of-plane polarization
   profile along the
   x-axis, for successive layers from the free surface (top) to the
   film/substrate interface (bottom), for a 5-unit cell film.}
\end{figure}

As the system is electrically neutral and
there are no free charges to screen the depolarizing field, a
ferroelectric state with polarization perpendicular to film surface
manifests through the formation of 180$^o$ stripe domains with alternating
polarity~\cite{tinte01,strei02,stac04,wu04,korn04}.  
Atomic-level simulations provide detailed local information of the 
polarization across the film. Figure 1 shows the cell-by-cell
out-of-plane polarization profile along the x-axis, for successive
layers from the free surface to the film/substrate interface, for a
5-unit cell film~\cite{com2}.
Positive polarization values correspond to polarization oriented 
towards the substrate.
Stripe domains aligned along the y-axis are formed.
This stripe domain pattern forms naturally and does not depend on the
initial configuration of the minimization procedure (a non-polarized or
uniformly polarized state). 
However, the domain period is imposed by the 
in-plane periodicity of the simulation cell.
It is clear from that figure that the polarization is not
uniform inside each nanodomain, and it increases with
the distance to the interfaces: free surface, film/substrate interface,
and domain walls. As a result, 180$^o$ domain walls in the film are
not as sharp as in bulk. 

\begin{figure}[tbp] 
\includegraphics[height=2.8in,width=3.35in]{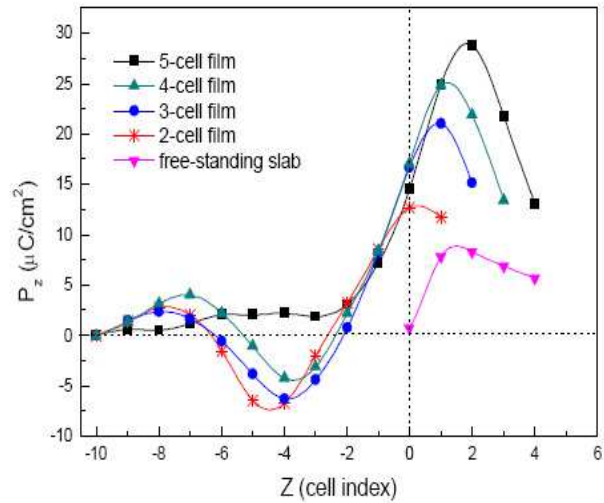}
   \caption{\label{fig2} (color online)
           Average layer-by-layer out-of-plane polarization profiles
          within a domain for films 2 to 5 unit cells thick on the ST
          substrate. 
	  For comparison, the profile for the free-standing 5-unit cell
          film is included.}
\end{figure}

Figure 2 shows the average layer-by-layer out-of-plane polarization 
profiles within a domain (P$_z$) for films 2 to 5 unit cells thick 
on the ST substrate. 
For comparison, the profile for the free-standing 5-unit cell film 
under the same in-plane condition is showed. 
In this Figure positive values of z indicate the PT film region 
while negative values indicate the ST substrate region. 
The profiles for the films on the substrate are qualitatively similar 
to the one of the free-standing slab.  
However, the polarization in the last case is substantially lower. 
At the interface (z=0), P$_z$ is only 1.2 $\mu$C/cm$^2$ in the 
free-standing slab and it increases to 14 $\mu$C/cm$^2$ 
when the substrate is present. 
For the inner layer, P$_z$ varies from 6 $\mu$C/cm$^2$ to 28 $\mu$C/cm$^2$ 
respectively. 
Since we take the same in-plane lattice parameter in both cases, 
the enhancement of the polarization is
due to the presence of the paraelectric substrate. 
The larger P$_z$ at the free surface (z=4) of the film on 
the substrate indicates that the interaction between them 
manifests throughout the film. 
The effect of the
film-substrate interaction is also reflected in the substrate region,
where an induced polarization is developed in the non-polar compound. 
In fact, Figure 2 shows that in the paraelectric substrate, a region of 
$\approx$ 3 unit cells near the interface polarizes in the same
direction as the film. These interface effects  can be understood from
the particular correlation of local dipole moments in perovskites. The
so called Lorentz field produces a significant mutual enhancement of the
dipoles along the direction of polarization. As a consequence,
strongly correlated chains of polarization tend to form.  As ST is a
highly polarizable substrate, the ionic displacements of the
ferroelectric film correlate with the ones of the substrate across the
coherent interface reinforcing the PT film polarization. As we will show
later, this correlation is crucial for the stabilization of a
ferroelectric state in films a few unit-cells thick.

The large increment of the film polarization produced by the presence of
the paraelectric substrate indicates the importance of selecting
appropriate boundary conditions to interprete correctly experimental
results. The ST substrate favors the ferroelectricity of PT films and,
as a consequence, it produces a decrement in the minimum thickness for
ferroelectricity, in comparison with the free-standing slab. In fact,
the minimum thickness decreases from 5 unit cells for the
free-standing slab~\cite{sepli051} to only 2 unit cells when the film is on the
ST substrate~\cite{coment}. The polarization profile of the 2-unit cell film
(Figure 2) shows that the film displays an average domain polarization
of $\approx$ 12 $\mu$C/cm$^2$.
It should be noted that our theoretical calculations are at
zero temperature while the experiments were performed above 120 K. This
means that the 2-unit cell film predicted to be ferroelectric at 0 K
might be expected to be paraelectric above 120 K. Moreover, the fit of
the critical temperature as a function of film thickness predicted that
the 2-unit cell thick film would be ferroelectric at low temperatures
~\cite{strei02}.  So, the substrate not only imposes an epitaxial strain on the
film  but generates a strong film-substrate coupling that plays an
active role in the ferroelectric behavior of the film.

Another interesting effect observed in the substrate region is that
the local polarization displays an unusual oscillatory behavior. The
polarization profiles shown in Figure 2 for the cases of 2 to 4-unit
cell films indicate that a region of the ST substrate near the interface
is polarized in the opposite direction than the film. 
The inversion of the induced polarization in the
paraelectric material, which appears 3 cells from the interface,
probably indicates that the depolarizing field has not been 
completely screened
by the domain configuration and overcomes the effect of the correlation
with the ferroelectric film. 
By extending the relaxed ST region up to 20 unit cells from the interface,
we observed that damped polarization oscillations continue beyond
the relaxed 10 cells of Figure 2. This fact, however, does not 
affect the polarization profiles of the PT films. 
A polarization invertion in the substrate region was obtained recently 
from estimations, under reasonable assumptions,
of the dipole moment per unit cell using atomic positions determined by
synchrotron x-ray scattering~\cite{fong05}. However, it was
presumed that this feature is a consequence of the assumptions made in the
calculations. 
We show here that this polarization inversion is indeed present 
when films are thin enough. 

\begin{figure}[tbp] 
 \includegraphics[height=2.8in,width=3.30in]{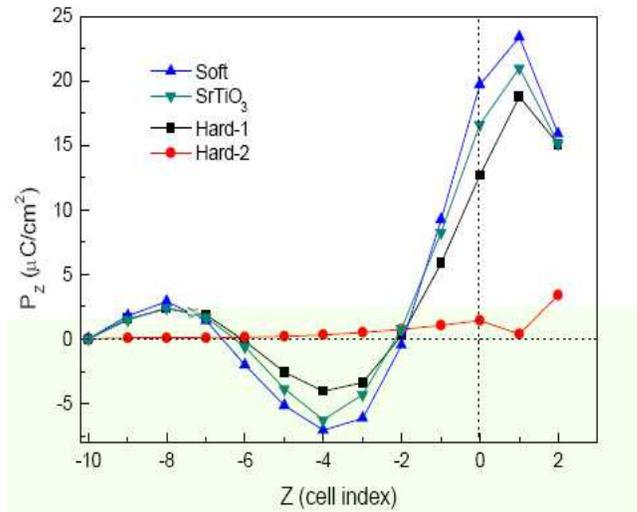}
   \caption{(color online) Average layer-by-layer out-of-plane
   polarization profiles for a 3-unit cell film on
   hypothetical ST-like paraelectric substrates with
   different polarizabities.}
  \label{fig3}
\end{figure}

The above results raise the question of how the polarizability of the
paraelectric substrate affects the minimum thickness for 
ferroelectricity in PT films. To answer this question we develop
several ST-like hypothetical compounds with the following
characteristics: they all have the same structure and lattice parameter
of ST, they are paraelectric, but they have a different soft-mode
frequency (i.e. different polarizability). We term these compounds Soft,
Hard-1, and Hard-2, with a soft-mode frequency of 19 cm$^{-1}$, 
181 cm$^{-1}$, and
243 cm$^{-1}$, respectively (the polar-mode frequency of the ST model 
is 77 cm$^{-1}$). We plot in Figure 3 the polarization
profile for a 3-unit cell film on the different hypothetical
paraelectric substrates. It is clear that the film polarization
increases when the substrate becomes softer (Soft), and decreases when
the substrate becomes harder (Hard-1). If the substrate is hard enough
(Hard-2), no domain structure or ferroelectricity is found.  So, not
only the polarization of the film but the minimum film thickness for
ferroelectricity depends on the dielectric properties of the
substrate. Figure 4 shows the minimum thickness for the stabilization of
a ferroelectric state as function of the substrate dielectric constant.
For a dielectric constant higher than the one of ST, the minimum film
thickness is 2 unit cells. This value increases when the substrate
hardens its dielectric response. In the limiting case of a
non-polarizable substrate, where the ions were held fixed at the
equilibrium cubic positions, a minimum thickness for ferroelectricity of
5 unit cells is obtained, in concordance with the critical thickness of
the free-standing slab. So, the critical thickness for the stabilization
of a spontaneous polarized state is not an intrinsic property of the
ferroelectric PT film but depends on the polarizability of the
paraelectric substrate.  
For the thinnest film with ferroelectric
properties observed in experiments, which is 3-unit cells on ST,
a cooperative phenomenon between film and substrate ions across 
the interface is responsible for the stabilization of the
ferroelectric state.
That is, substrate atoms are part of the
correlated region and contribute to the minimum correlation length
necessary for ferroelectricity.

\begin{figure}[tbp] 
 \includegraphics[height=2.8in,width=3.30in]{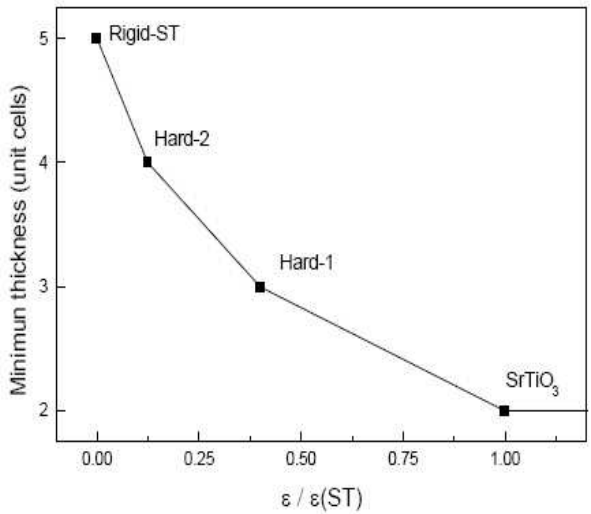}
   \caption{Minimum thickness for the stabilization of
   a ferroelectric state as function of the substrate dielectric constant.}
  \label{fig4}
\end{figure}

The effects of the ferroelectric-parelectric coupling 
is even stronger when the PT film is coherently strained 
between two ST layers. 
In fact, we simulate PT ultrathin films of 
different thickness between two ST layers of 10 unit cells thick. 
The results indicate that, under such a boundary condition, 
even a single PbO plane polarizes in a stripe domain configuration, 
with a local polarization of 7 $\mu$C/cm$^2$.  
This polarization is favored by the coupling to the
surrounding highly polarizable media, which manifests by the extension
of the polarized region over 5 unit cells.  Interestingly, the
paraelectric regions at both sides of the ferroelectric layer also
display polarization oscillations, which might produce unexpected
effects in superlattices with PT layers a few unit cells thick~\cite{dawb052}.

\acknowledgments
This work was supported by Agencia Nacional de Promoci\'on Cient\'{\i}fica y 
Tecnol\'ogica, and
Consejo Nacional de Investigaciones Cientif\'{\i}cas y T\'ecnicas (Argentina).
M.G.S. thanks support from Consejo de Investigaciones de la Universidad Nacional de
Rosario. \\ \\


\begin{references}
\bibitem{com2} We note that, as a consequence of LDA, we expect polarization
values underestimated with respect to the experimental ones.
\bibitem{coment} The one unit-cell thick film does not display a polar state, no
domain structure or ferroelectricity is found. In this film, the surface
cells develop a small inward polarization due to the relaxation and
rumpling of the surface layers.
\end{references}
\end{document}